\documentclass[12pt,preprint]{aastex}

\newcommand{\kms}{\mathrm{km\,s}^{-1}}
\newcommand{\ms}{\mathrm{m\,s}^{-1}}

\shorttitle{Relativity in spectroscopic binaries}
\shortauthors{Zucker \& Alexander}

\begin{document}

\title{Spectroscopic Binary Mass Determination using
Relativity}

\author{Shay Zucker\altaffilmark{1,2} and Tal Alexander\altaffilmark{2,3}} 
\altaffiltext{1}{Dept. of Geophysics \& Planetary Sciences, 
	Raymond and Beverly Sackler Faculty of Exact Sciences, Tel
	Aviv University, Tel Aviv 69978, Israel}
\altaffiltext{2}{Faculty of Physics, Weizmann Institute of Science, 
	PO Box 26, Rehovot 76100, Israel}
\altaffiltext{3}{The William Z. \& Eda Bess Novick career development chair}

\email{shayz@post.tau.ac.il,
	Tal.Alexander@weizmann.ac.il}

\begin{abstract}
High-precision radial-velocity techniques, which enabled the detection
of extrasolar planets are now sensitive to relativistic effects in the
data of spectroscopic binary stars (SBs). We show how these effects
can be used to derive the absolute masses of the components of
eclipsing single-lined SBs and double-lined SBs from Doppler
measurements alone. High-precision stellar spectroscopy can thus
substantially increase the number of measured stellar masses, thereby
improving the mass-radius and mass-luminosity calibrations.
\end{abstract}


\section{Introduction}

Precise radial-velocity (RV) measurements, with long-term precisions
of a few meters per second, are now routinely obtained by several
telescopes around the world. The most notable scientific achievement
of precise RV measurements has been the detection of planets orbiting
solar-type stars \citep{MayQue1995,MarBut1996}. 
In this {\it Letter} we suggest another application of high-precision
RVs, namely, the detection of relativistic effects in the Doppler
shifts of close spectroscopic binary stars (SBs). \citet{KopOze1999}
have already detailed the relativistic effects one expects to find in
the Doppler measurements of binary stars. Here we focus on the effects
that we expect to measure in SBs, and identify the additional
information they provide.

The typical velocities of components of close binary stars can be as
high as $150\,\kms$, $\beta \equiv v/c \sim {\cal O}(10^{-4})$. The
classical Doppler shift formula predicts a relative wavelength shift
$\Delta \lambda/\lambda$ of order $\beta$. The next order corrections
are of order $\beta^2 \sim {\cal O}(10^{-8})$, which translate to
${\cal O}(1\,\ms)$. Terms of order $\beta^3$ are beyond foreseen
technical capabilities. We thus limit our analysis to ${\cal
O}(\beta^2)$ effects --- the transverse Doppler shift (time dilation)
and the gravitational redshift. In principle, long term monitoring may
reveal higher-order secular terms, such as the relativistic periastron
shift or period decay through gravitational-wave radiation, but here
we focus only on the periodic effects. These can be detected during
relatively short observing runs, assuming the RV measurements are
precise enough.

Recently, \citet{Zucetal2006} have shown that the very same effects
should be detectable in the stellar orbits around the black hole in
the Galactic Center, after a decade of observations. The context we
examine here is different, and the information that can be extracted
may contribute to the statistics of close binary orbits.

\section{Single-lined spectroscopic binary}

The Keplerian RV curve of a single-lined spectroscopic binary (SB1)
can be presented as: 
\begin{equation}
\label{SB1simple}
V_{R1} = K_1 \cos \omega \cos \nu - K_1 \sin \omega \sin \nu + e K_1
\cos \omega + V_{R0} \ ,
\end{equation}
where $K_1$ is the primary star RV semi-amplitude, $\omega$ is the
argument of periastron, $e$ is the eccentricity, $V_{R0}$ is the
center-of-mass RV, and $\nu$ is the time-dependent true anomaly. The
customary procedure to solve an SB1 is to fit this orbital model to
the observed RV data. This fit is achieved through some optimization
algorithm that scans the $(P,T,e)$ space (period, periastron time and
eccentricity). For each trial set of values for these three parameters
the algorithm produces the corresponding $\nu(t)$ and then solves
analytically for $K_{C1}$($=K_1\cos\omega$),
$K_{S1}$($=-K_1\sin\omega$), and $V_{R0}$, which appear linearly in
the expression for $V_{R1}$.

In order to incorporate relativity into the observed RV curve of an
SB1 we can use the models developed for analyzing binary pulsar timing
data.  \citet{TayWei1989} present a detailed timing model of a
relativistic binary pulsar, based on the relativistic celestial
mechanics developed by \citet{DamDer1986}. Besides the transverse
Doppler shift and the gravitational redshift, relativity also
introduces the Shapiro delay, periastron advance, and period decay
through gravitational wave radiation. The formulae for the pulse delay
in \citet{TayWei1989} can be transformed to the RV domain by taking
their time derivative. This calculation shows that the only terms of
order $\beta^2$ are the gravitational redshift and the transverse
Doppler shift, corresponding to the so-called 'Einstein delay' in the
pulsar timing model.

We now derive the two terms in a more didactic, albeit less rigorous
fashion.  In the center-of-mass frame, we can use energy conservation
in the classic Keplerian solution to relate the transverse Doppler
term to the radius vector of the observed component, $r_1$:
\begin{equation}
\label{energycons}
\frac{1}{2}\beta^2 = \frac{1}{c^2} \frac{Gm_2^3}{\left(m_1+m_2\right)^2}\left(\frac{1}{r_1}-\frac{1}{2a_1}\right)
\end{equation}
where $a_1$ is the orbital semi-major axis of the primary orbit, and
$c$ is the speed of light.  After some algebraic manipulation, we can
estimate the corresponding modification to the measured RV from the
transverse Doppler effect:
\begin{equation}
\label{transdopp}
\Delta V_{\mathrm{TD}} = \frac{K_1^2}{c\sin^2i}\left(1+e\cos\nu-\frac{1-e^2}{2}\right)
\end{equation}
where $i$ is the orbital inclination.

The gravitational redshift caused by the potential of the secondary
component is inversely proportional to the separation of the two
components:
\begin{equation}
\label{grterm}
\frac{\Delta\lambda}{\lambda} = \frac{G m_2}{c^2\left(r_1+r_2\right)}
\end{equation}
and the corresponding RV modification term from the gravitational
redshift is:
\begin{equation}
\label{gravredshift}
\Delta V_{\mathrm{GR}} = \frac{K_1\left(K_1+K_2\right)}{c\sin^2i}\left(1+e\cos\nu\right)
\end{equation}
After transformation to the observer frame, an additional term
appears, related to the center-of-mass motion
\begin{equation}
\label{comtransdopp}
\frac{V_{R0}}{c}K_1\left(\cos(\nu+\omega)+e\cos\omega\right)+\frac{V_0^2}{2c}\ ,
\end{equation}
where $V_0$ is the magnitude of the full center-of-mass velocity
vector.

In total, relativity adds the following ${\cal O}(\beta^2)$ term to
the measured $V_{R1}$: 
\begin{eqnarray}
\label{allterms}
\Delta V_{R1} &= &\frac{K_1}{c} \frac{1}{\sin^2i} \left [
e (2 K_1 + K_2) \cos\nu + (2 K_1 + K_2) - \frac{1-e^2}{2} K_1 \right ] +  \\
&&\frac{V_{R0}}{c}K_1 \left(\cos\omega\cos\nu-\sin\omega\sin\nu+e\cos\omega\right) +
\frac{V_0^2}{2c} \nonumber
\end{eqnarray}
The resulting expression for the modified measured RV, $V_{R1}'$, can
now be easily simplified to:
\begin{equation}
\label{modifiedrv}
V_{R1}' = K_{C1}'\cos \nu + K_{S1}'\sin \nu + e K_{C1}' + V_{R0}' \ ,
\end{equation}
by collecting together the constant terms, the terms proportional to
$\cos\nu$, and those proportional to $\sin\nu$, and introducing the
modified linear elements:
\begin{mathletters}
\label{modifiedelementsSB1}
\begin{eqnarray}
K_{S1}' &= &-K_1 \left( 1+\frac{V_{R0}}{c} \right ) \sin \omega \label{SB1KS1} \\
K_{C1}' &= &K_1 \left [ \left ( 1+\frac{V_{R0}}{c} \right ) \cos \omega + \frac{e}{\sin^2i} \frac{2K_1+K_2}{c} \right ] \label{SB1KC1} \\
V_{R0}' &= &V_{R0} + \frac{1-e^2}{\sin^2i}\frac{K_1}{c}\left(\frac{3}{2}K_1+K_2\right)+\frac{V_0^2}{2c} \label{SB1VR0}
\end{eqnarray}
\end{mathletters}

Equations \ref{SB1simple} and \ref{modifiedrv} share exactly the same
structure, and thus we can still apply the same fit
procedure. However, the linear elements are more difficult to
interpret now. The three quantities $K_{S1}'$, $K_{C1}'$, and
$V_{R0}'$ depend on the six elements $K_1$, $K_2$, $\omega$, $\sin i$,
$V_{R0}$, and $V_0$. Thus, Equations \ref{modifiedelementsSB1} are
under-determined and we cannot completely infer the six elements
above, unless some additional independent information is available, or
further assumptions are introduced.

Such independent information may be available through precise
photometry of eccentric eclipsing binaries. There, the shapes and
widths of the eclipses as well as the phase differences between
primary and secondary eclipses can be used to estimate $\omega$ and
$\sin i$. In this case, we may derive $K_2$ -- the RV amplitude of the
secondary:
\begin{equation}
\label{K2expression}
K_2 = \frac{2K_{S1}'}{\sin\omega} -
c\frac{\sin^2i}{e}\left(\cos\omega+\frac{K_{C1}'}{K_{S1}'}\sin\omega\right) \ .
\end{equation}
In the above equation we neglected $V_{R0}$, as it contributes only
higher order terms. By obtaining $K_2$ we effectively turn the binary
into a double-lined spectroscopic binary (SB2), in which both $K_1$
and $K_2$ are measured. Together with the known inclination, we then
obtain full knowledge of the component masses.

Curiously, another result of including the relativistic terms is that
an eccentric binary should always display an apparent RV signature,
even in the extreme case where the inclination is exactly zero and the
orbit is observed face on. Since $K_j = 2 \pi a_j \sin i/(P
\sqrt{1-e^2})$, $\sin i$ will cancel out in Equation
\ref{SB1KC1}. Then $K_{C1}'$ will be finite, while $K_{S1}' = 0$, and
the apparent argument of periastron will be exactly $0\degr$ or
$180\degr$. Thus, all the RV planet candidates whose arguments of
periastra are close to these values, can in principle be binary stars
observed exactly face on. Nevertheless, such small values of the
inclination are extremely rare and this possibility is not realistic.

Equation \ref{SB1VR0} does not contribute any new useful information,
since systematic effects in the measurement process such as spectral
template mismatch are probably larger than the relativistic
effects. In addition, the spectra are subject to gravitational
redshift by the potential of the emitting star itself, and the typical
uncertainties regarding its mass and radius are also larger than the
effects we discuss here.

\section{Double-lined spectroscopic binary}

In the case of a Keplerian SB2, there are two sets of measured
RVs. The two sets of RVs share the same fundamental orbital elements
($P$, $T$, $e$, $\omega$, and $V_{R0}$), and the only difference
between them is their amplitudes $K_1$ and $-K_2$. The common
procedure is to scan the space of the four parameters $(P,T,e,\omega)$
and then solve analytically for the three linear elements $K_1$,
$K_2$, and $V_{R0}$. However, when we incorporate the relativistic
corrections above, we see that we now have two RV curves with two
different sets of derived amplitudes, arguments of periastra, and
center-of-mass velocities. The two RV curves still share the same
period, periastron time and eccentricity. We then have the following
equations, corresponding to Equations \ref{SB1KS1} and \ref{SB1KC1}:
\begin{mathletters}
\label{modifiedelementsSB2}
\begin{eqnarray}
K_{S1}' &= &-K_1 \left( 1+\frac{V_{R0}}{c} \right ) \sin \omega \label{SB2KS1} \\
K_{C1}' &= &K_1 \left [ \left ( 1+\frac{V_{R0}}{c} \right ) \cos \omega + \frac{e}{\sin^2i} \frac{2K_1+K_2}{c} \right ] \label{SB2KC1} \\
K_{S2}' &= &-K_2 \left( 1+\frac{V_{R0}}{c} \right ) \sin \omega \label{SB2KS2} \\
K_{C2}' &= &K_2 \left [ \left ( 1+\frac{V_{R0}}{c} \right ) \cos \omega - \frac{e}{\sin^2i} \frac{K_1+2K_2}{c} \right ] \label{SB2KC2}
\end{eqnarray}
\end{mathletters}

$V_{R0}$ appears in the four equations always divided by the speed of
light. Thus we can safely use its approximate derived value, from
either set of measured velocities , since the discrepancy will be only
${\cal O}(\beta^3)$.  We are left with four equations with four
unknowns: $K_1$, $K_2$, $\omega$, and $\sin i$. Solution of this set
of equations will yield a more accurate estimate of the first three
unknowns, but more importantly, it will yield an estimate of $\sin
i$. Retaining only leading order terms, we can arrive at the following
solution:
\begin{equation}
\label{siniexpression}
\sin^2i = \frac{3 e}{\omega_2'-\omega_1'}\frac{K_{S2}'+K_{S1}'}{c} \ ,
\end{equation}
where
\begin{equation}
\omega_j' = -\arctan\left(\frac{K_{S_j}'}{K_{C_j}'}\right) \ .
\end{equation}

Thus, we see that relativity causes the apparent arguments of
periastra to differ. Note that classically, there is no way to
estimate $\sin i$ from pure RV data alone. The value of $\sin i$ is
usually obtained only from the analysis of SBs that are eclipsing or
astrometric binaries.

\section{Discussion}
We present here an approach to extract more information from RV data
of a spectroscopic binary, based on the inclusion of relativistic
effects. The practical potential lies in the time-dependent parts of
the relativistic terms which are closely linked to the variation of
the distance between the binary components. Thus, these terms are
especially useful for orbits that are eccentric enough.

The approach is mainly useful in the case of SB2s, but precise
photometry or astrometry may add the required information to utilize
relativity in SB1s as well. Precise photometry space missions like
MOST \citep{Waletal2003}, CoRoT \citep{Bag2003}, and Kepler
\citep{Basetal2005} may be able to provide precise enough measurements
of $\omega$ and the inclination of eclipsing binaries from the
analysis of their light curves.

In any case, it is essential that the data quality be high enough to
be sensitive to variations of the required order, namely one meter per
second or less. Currently, the best precision is obtained by the ESO
HARPS fiber-fed echelle spectrograph, where the RV error can be as low
as $1\,\ms$ in certain cases and maybe even less
\citep{Lovetal2005}. In the future, much better precisions can be
hoped for, on instruments designed for the ``extremely large
telescopes'' \citep[e.g.,][]{Pasetal2006}.

\begin{figure}
\plotone{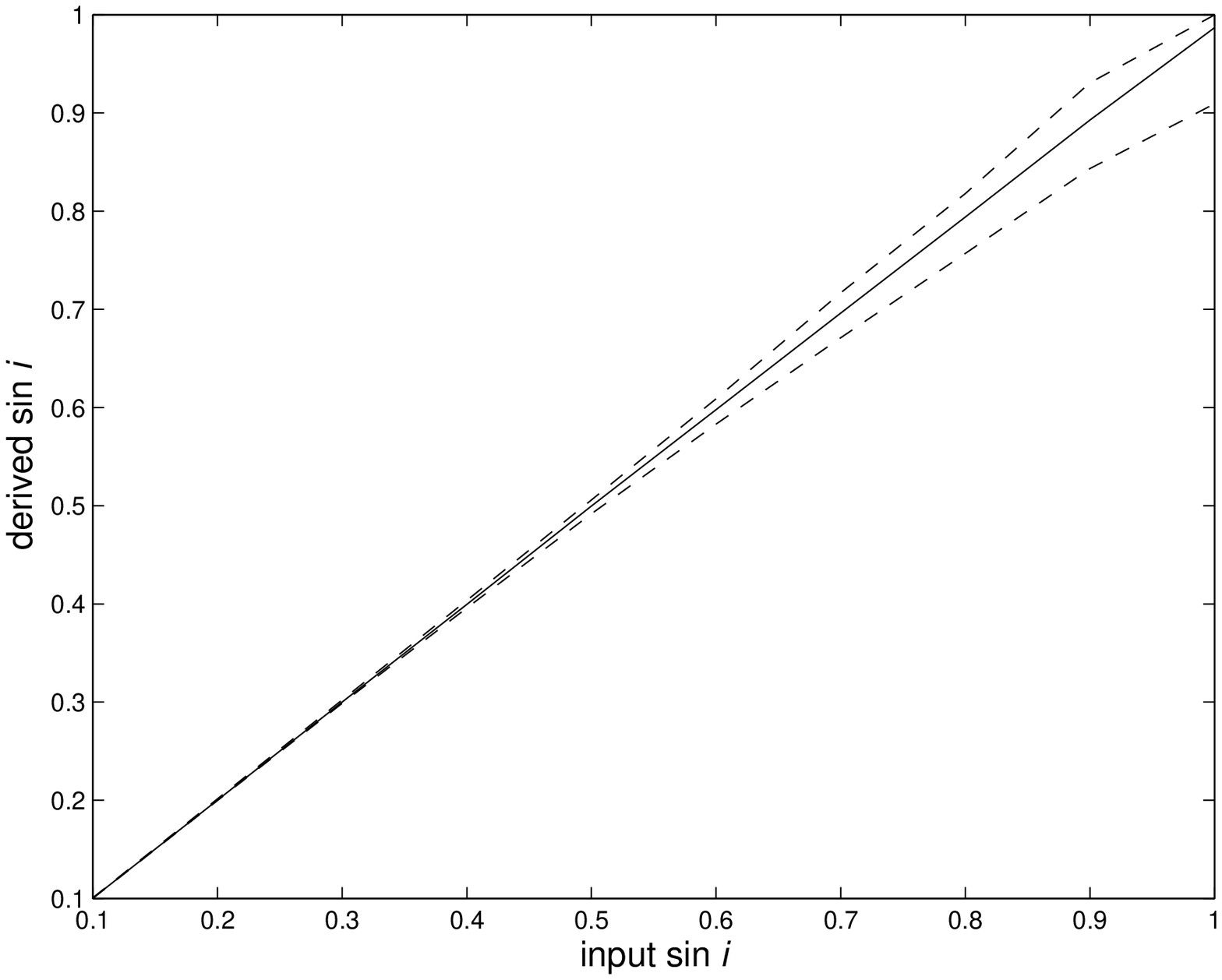}
\caption{The results of simulations of 12\,Boo including relativistic
effects, assuming RV precision of $1$\,m\,s$^{-1}$. The plot shows
percentiles of the derived $\sin i$ against the input simulated
value. The dashed lines represent the first and third quartile, and
the solid line is the median.}
\label{simsini}
\end{figure}

To demonstrate the relevance of the suggested approach in real-life
cases we chose to examine the SB2 12\,Boo. Recently,
\citet{TomFek2006} published a precise solution of 12\,Boo based on
RVs obtained at the $2.1$-m telescope at the McDonald Observatory and
at the Coud\'{e} feed telescope at Kitt Peak National Observatory,
with RV precisions of $0.1$--$0.2$\,km\,s$^{-1}$.  The system has a
period of $9.6$ days, eccentricity $0.2$, and both RV semi-amplitudes
are close to $70$\,km\,s$^{-1}$. These orbital parameters translate to
an expected relativistic amplitude variation of about
$10$\,m\,s$^{-1}$ (Eq. \ref{allterms}).  Furthermore, with a
declination of $+25\degr$, the star is observable by HARPS. Its
brightness (5th magnitude) and spectral type (F9IV) make it fairly
reasonable to expect a precision of $1$\,m\,s$^{-1}$ with HARPS.  We
used the available $24$ RVs from \citet{TomFek2006} and augmented them
with only $3$ simulated HARPS measurements (assuming errors of
$1$\,m\,s$^{-1}$) , including the relativistic effects. For each
assumed value of $\sin i$ we produced $1000$ sets of simulated
measurements and solved for the orbital elements, using Equation
\ref{siniexpression} to estimate $\sin i$. Figure \ref{simsini} shows
the median of the derived values in solid line, and the $25\%$ and
$75\%$ percentiles in dashed lines.  The figure demonstrates that with
reasonable efforts, $\sin i$ can be measured satisfactorily. In the
worst case where $\sin i=1$, the standard deviation of the derived
inclination is $0.14$ and a few more precise measurements can reduce
this value significantly. An additional advantage of this test case is
that the inclination of 12\,Boo has already been measured by
interferometry and is known to be $108\degr$ \citep{Bodetal2005}. Thus
if this test is performed, the derived $\sin i$ can be compared to the
known value.

In real observations, more than three precise measurements may be
needed in order to account for differences in zero points between
instruments. Furthermore, The results depend crucially on the
precision of $\omega_1'$ and $\omega_2'$, and a relative error of
$\epsilon$ in $\omega_2'-\omega_1'$ will be translated to a relative
error of $1.5\epsilon$ in the absolute masses.  The few precise
measurements can be scheduled to optimize the precision of those two
elements \citep[e.g.][]{For2006}.

Currently, precisions of a few meters per second are still difficult
to obtain and besides using the best instruments available, there are
also several limitations imposed by the star itself. Thus, early-type
stars or rapid rotators, where the spectral lines are significantly
broadened, do not lend themselves easily to high-precision RV
measurements. Stellar oscillations and star spots are also a concern
as they can cause apparent RV modulation.  In addition, analyzing SB2s
with the same level of precision as SB1s has not been easy until
recently, when TODCOR \citep{ZucMaz1994} was applied successfully to
high-precision spectra by several teams
\citep{Zucetal2004,Udretal2004,Kon2005}.

The effects we have examined are most useful when the orbits are
eccentric and the RV amplitudes are large enough. Large RV amplitudes
are usually typical to close binaries, which are expected to have
undergone orbital circularization and usually have vanishing
eccentricities. However, eccentricity somewhat increases the RV
amplitude, and even relatively wide binaries, with high enough
eccentricities, can display quite large RV amplitudes.

Care must be taken to model correctly any other effects of order
$\beta^2$ that might contaminate the data. One such effect is the
light-travel-time effect. This effect can be easily approximated to
the relevant order by adding the following term to $V_{R1}$ (and a
corresponding one to $V_{R2}$):
\begin{equation}
\Delta V_{\mathrm{LT}} = \frac{K_1^2}{c}\sin^2\left(\nu+\omega\right)\left(1+e\cos\nu\right)
\end{equation}

An effect which should be analyzed carefully is the tidal distortion
of the stellar components, in particular close to periastron. This
distortion may affect the spectral lines, introducing line asymmetry,
which can bias the estimated Doppler shift. RV extrasolar planet
surveys use the line-bisector analysis \citep[e.g.,][]{Queetal2001} to
quantify such time-dependent asymmetries. Further development of this
technique may be the key to disentangle the tidal distortion and the
relativistic effects.


One important application of the proposed method is to calibrate the
low-mass end of the mass-luminosity relation, to better understand the
stellar-substellar borderline. This mass regime is still poorly
constrained, since low mass SB2s are quite rare due to the special
photometric, spectroscopic and geometric requirements
\citep{Rib2006}. Large efforts are in progress to obtain accurate
stellar masses in this regime, including adaptive optics,
interferometry and in the future space interferometry
\citep[e.g.,][]{Henetal2005}. We propose a new, relatively accessible
tool to accomplish this goal, where the only requirements are
spectroscopic.  Precise RVs for low-mass SB2s were already measured by
\citet{Deletal1999}.  Using the method presented here, their absolute
masses may be derived with a relatively small observational effort.
No other method exists yet to derive this information purely from RV
measurements.

\acknowledgments
We are grateful to Tsevi Mazeh for very useful and stimulating
discussions.  T.A. is supported by Minerva grant 8563 and a New
Faculty grant by Sir H. Djangoly, CBE, of London, UK.

\clearpage

\begin{thebibliography}{}

\bibitem[Baglin(2003)]{Bag2003}
	Baglin, A. 2003, Adv. Space Res., 31, 345


\bibitem[Basri, Borucki \& Koch(2005)]{Basetal2005}
	Basri, G., Borucki, W. J., \& Koch, D. 2005, \nar, 49, 478

\bibitem[Boden et al.(2005)]{Bodetal2005}
	Boden, A. F., Torres, G., \& Hummel, C. A. 2005, \apj, 627, 464

\bibitem[Damour \& Deruelle(1986)]{DamDer1986}
	Damour, T., \& Deruelle, N. 1986, Ann. Inst. H. Poincar\'{e}
	(Physique Th\'{e}orique), 44, 263

\bibitem[Delfosse et al.(1999)]{Deletal1999}
	Delfosse, X., Forveille, T., Beuzit, J.-L., Udry, S., Mayor,
	M., \& Perrier, C. 1999, \aap, 344, 897


\bibitem[Ford (2006)]{For2006}
        Ford, E. B. 2006, \aj, submitted (astro-ph/0412703)


\bibitem[Henry et al.(2005)]{Henetal2005}
	Henry, T. J., et al. 2005, \baas, 37, 1356

\bibitem[Konacki(2005)]{Kon2005}
	Konacki, M. 2005, \apj, 626, 431


\bibitem[Kopeikin \& Ozernoy(1999)]{KopOze1999}
	Kopeikin, S. M., \& Ozernoy, L. M. 1999, \apj, 523, 771

\bibitem[Lovis et al.(2005)]{Lovetal2005}
	Lovis, C., et al. 2005, \aap, 437, 1121


\bibitem[Marcy \& Butler(1996)]{MarBut1996}
	Marcy, G. W., \& Butler, R. P. 1996, \apjl, 464, L147

\bibitem[Mayor \& Queloz(1995)]{MayQue1995}
	Mayor, M., \ Queloz, D. 1995, \nat, 378, 355


\bibitem[Pasquini et al.(2006)]{Pasetal2006}
	Pasquini, L., et al. 2006, in IAU Symp. 232, The Scientific
	Requirements for Extremely Large Telescopes, ed. P. Whitelock,
	B. Leibundgut, \& M. Dennefeld (Cambridge: Cambridge
	Univ. Press), in press

\bibitem[Queloz et al.(2001)]{Queetal2001}
	Queloz, D., et al. 2001, \aap, 379, 279

\bibitem[Ribas(2006)]{Rib2006}
	Ribas, I. 2006, \apss, in press (astro-ph/0511431)

\bibitem[Taylor \& Weisberg(1989)]{TayWei1989}
	Taylor, J. H., \& Weisberg, J. M. 1989, \apj, 345, 434

\bibitem[Tomkin \& Fekel(2006)]{TomFek2006}
	Tomkin, J., \& Fekel, F. C. 2006, \aj, 131, 2652

\bibitem[Udry et al.(2004)]{Udretal2004}
	Udry, S., Eggenberger, A., Mayor, M., Mazeh, T. \& Zucker,
	S. 2004, \rmxaa, 21, 207

\bibitem[Walker et al.(2003)]{Waletal2003}
	Walker, G., et al. 2003, \pasp, 115, 1023

\bibitem[Zucker \& Mazeh(1994)]{ZucMaz1994}
	Zucker, S., \& Mazeh, T. 1994, \apj, 420, 806

\bibitem[Zucker et al.(2006)]{Zucetal2006}
	Zucker, S., Alexander, T., Gillessen, S., Eisenhauer, F., \&
	Genzel, R. 2006, \apjl, 639, L21

\bibitem[Zucker et al.(2004)]{Zucetal2004}
	Zucker, S., Mazeh, T., Santos, N. C., Udry, S., \& Mayor,
	M. 2004, \aap, 426, 695

\end{thebibliography}
\end{document}